\documentclass[namedreferences]{solarphysics}
\usepackage[optionalrh,solaromanenum]{spr-sola-addons} 
\usepackage{graphicx}                    
\usepackage{color}                       
\usepackage{url}                         

\newcommand{\etal}{{\it et al.}}
\newcommand{\insitu}{{\it in-situ }}
\newcommand{\kms}{{$\mathrm{km\,s^{-1}}$}}

\begin{document}
\begin{article}
\begin{opening}

\title{Relationship Between Solar Wind Speed and Coronal Magnetic Field Properties}

\date{\today}
\author{Ken'ichi~\surname{Fujiki}$^{1}$\sep
       Munetoshi~\surname{Tokumaru}$^{1}$\sep
       Tomoya~\surname{Iju}$^{1}$\sep
       Kazuyuki~\surname{Hakamada}$^{2}$\sep
       Masayoshi~\surname{Kojima}$^{1}$
       }

%
\runningauthor{K. Fujiki \etal}
\runningtitle{Solar Wind Velocity and Coronal Magnetic Field }

%

\institute{
$^{1}$ Solar-Terrestrial Environment Laboratory, Nagoya University, 3-13 Honohara, Toyokawa, Aichi 442-8507 Japan 
email: \url{fujiki@stelab.nagoya-u.ac.jp}\\ 
$^{2}$ Department of Natural Science and Mathematics, Chubu University, 1200 Matsumoto-cho, Kasugai, Aichi, 487-8501 Japan\\
}

\begin{abstract}
We have studied the relationship between the solar-wind speed $[V]$ and the coronal magnetic-field properties (a flux expansion factor [$f$] and photospheric magnetic-field strength [$B_{\mathrm{S}}$]) at all latitudes using data of interplanetary scintillation and solar magnetic field obtained for 24 years from 1986 to 2009. Using a cross-correlation analyses, we verified that $V$ is inversely proportional to $f$ and found that $V$ tends to increase with $B_{\mathrm{S}}$ if $f$ is the same. As a consequence, we find that $V$ has extremely good linear correlation with $B_{\mathrm{S}}/f$. However, this linear relation of $V$ and $B_{\mathrm{S}}/f$ cannot be used for predicting the solar-wind velocity without information on the solar-wind mass flux. We discuss why the inverse relation between $V$ and $f$ has been successfully used for solar-wind velocity prediction, even though it does not explicitly include the mass flux and magnetic-field strength, which are important physical parameters for solar-wind acceleration. 
\end{abstract}


\end{opening}

%
\section{Introduction}  \label{Introduction}
\citeauthor{Wang1990ApJ} (\citeyear{Wang1990ApJ,Wang1991ApJ}) showed that the solar-wind velocity $[V]$ correlated inversely with the expansion factor $[f]$ of a magnetic-flux tube in the corona, and this inverse relation has been improved by \citeauthor{Arge2000JGR} (\citeyear{Arge2000JGR})  for solar-wind prediction as $V=267.5+410/f^{2/5}$ (hereafter the WSA equation). \citeauthor{Wang1996Sci} (\citeyear{Wang1996Sci}) explained this inverse relation with observational data by the difference of $f$-dependences of the total-energy flux density at the sun and the ion flux density at the sun: although both parameters increase with $f$, the increase of the flux density is steeper than that of the total energy flux. 

\citeauthor{Fisk1999JGR} (\citeyear{Fisk1999JGR}) and \citeauthor{Fisk2003JGRA} (\citeyear{Fisk2003JGRA}) proposed the solar-wind acceleration model (hereafter the FSK equation),
\begin{equation} \label{eqFSK}
\frac{V_{\mathrm{f}}^2}{2}= \frac{P_\mathrm{i}}{\rho_\mathrm{i} u_\mathrm{i}} - \frac{\mathrm{G} \mathrm{M}_\odot}{\mathrm{R}_\odot},
\end{equation}
where $V_\mathrm{f}$, $P_\mathrm{i}$, $\rho_\mathrm{i} u_\mathrm{i}$, $\mathrm{G}$, $\mathrm{M}_\odot$, and $\mathrm{R}_\odot$ are the final solar-wind speed (\textit{e.g.} at 1 AU), the Poynting flux at coronal base, solar-wind mass flux at coronal base, the gravitational constant, the solar mass, and the solar radius, respectively.  The FSK equation indicates the final solar-wind velocity is determined by two parameters, mass flux released from the loops by reconnection with open field lines and the Poynting flux at coronal base. 

\citeauthor{Suzuki2002ApJ} (\citeyear{Suzuki2002ApJ}, \citeyear{Suzuki2006ApJ}) introduced the dissipation mechanism of MHD waves through shock process to the additional acceleration process of the solar wind, and proposed the following equation (hereafter the SZK equation),
\begin{equation} \label{eqSZK}
V^2_\mathrm{1AU}=2 \left[-\frac{\mathrm{R}^2_\odot}{4\pi \left( \rho v r^2 \right)_\mathrm{1AU}}\frac{B_{\mathrm{S}}}{f} \langle \delta B_\perp \delta v_\perp \rangle_\mathrm{S} + \frac{\gamma}{\gamma-1}\mathrm{R} T_\mathrm{C} - \frac{\mathrm{G} \mathrm{M}_\odot}{\mathrm{R}_\odot} \right], 
\end{equation}
where $\mathrm{R}$, $\gamma$, and $B_\mathrm{S}$ are, respectively, the gas constant, ratio of specific heats, and photospheric magnetic-field strength. $-\langle \delta B_\perp \delta v_\perp \rangle_\mathrm{S}$ indicates the Poynting flux of the sheared magnetic field, corresponding to Alfve\'n wave ($-\langle \delta B_\perp \delta v_\perp \rangle_\mathrm{S}>0$ for an outgoing component). Therefore, the SZK equation is a hybrid of WSA and FSK models. These three equations have the following similarities: i) in WSA and SZK equations, $V$ is proportional to $1/f$, ii) in the WSA and SZK equations, the $f$-dependent term works as an additional acceleration, iii) in the FSK and SZK equations, the Poynting flux is used as acceleration energy. iv) Another difference is the $B_{\mathrm{S}}$-dependence of $V$; although the SZK equation includes $B_{\mathrm{S}}$ explicitly, the WSA and FSK equations are independent of $B_{\mathrm{S}}$. 

The WSA equation has been widely and successfully used for solar-wind velocity prediction because of its simplicity. However, the physical parameters necessary for the solar-wind acceleration process, such as mass flux and energy are dropped. In order to elucidate the solar-wind acceleration process, we must make a model which includes all necessary corona physical parameters explicitly, such as the FSK and SZK equations. From this viewpoint, $B_{\mathrm{S}}$ will be one of the key parameters. \citeauthor{Wang2010ApJL} (\citeyear{Wang2010ApJL}) explained why the WSA equation can drop $B_{\mathrm{S}}$ and the mass flux: both the mass flux and energy flux at the Sun tend to increase approximately linearly with increasing $B_{\mathrm{S}}$, but $B_{\mathrm{S}}$ has no systematic relation with $V$ at the Earth: accordingly, the independence of $V$ on $B_{\mathrm{S}}$ causes no inclusion of the mass flux and the energy flux in the equation. This means that if we found a $B_{\mathrm{S}}$--$V$ relation, then, necessarily $B_{\mathrm{S}}$-related parameters of the mass flux and the energy flux should be included. First of all, we will verify the WSA equation in Section \ref{flux}. In Sections \ref{Bphoto} and \ref{Bf} we will examine the $B_{\mathrm{S}}$--$V$ relation and demonstrate that the SZK equation is the model preferable over the others from the viewpoint of solar-wind physics. In Section \ref{discussion}, however, we will show that the WSA equation works better than the SZK equation for solar-wind prediction, and explain the reason why $B_{\mathrm{S}}$ and the mass flux can be eliminated from the SZK equation.

This study was began by \citeauthor{Hirano2003AGUFM} (\citeyear{Hirano2003AGUFM}) with data obtained from 1995 to 1996. Analyses were made of large-scale coronal holes (CHs), which consist of a polar CH, a distinct equatorial- and mid-latitude CH, and a low-latitude polar CH extension. It was revealed that $V$ and $B_{\mathrm{S}}/f$ have a remarkably high correlation, with a correlation coefficient of 0.88 (\citeauthor{Kojima2004ASSL}, \citeyear{Kojima2004ASSL}, \citeyear{Kojima2007ASPC}). Based on their study, here we extend their analyses of data obtained for 24 years from 1986 to 2009, including two solar minimum phases. In this study, we were particularly interested in comparing the two solar minimum phases of 1994\,--\,1996 and 2007\,--\,2009, because the latter minimum phase showed peculiar aspects compared with the previous one (e.g., \citeauthor{McComas2008GeoRL}, \citeyear{McComas2008GeoRL}; \citeauthor{Tokumaru2009GeoRL}, \citeyear{Tokumaru2009GeoRL}; \citeauthor{vonSteiger2011JGRA}, \citeyear{vonSteiger2011JGRA}; \citeauthor{Wang2013ApJ}, \citeyear{Wang2013ApJ}). Different kinds of CHs, such as a polar CH, a middle latitude CH, and a CH associated with an active region, have different magnetic-field properties. However, most CHs are at middle and high latitudes where spacecraft cannot approach, with the exception of \textit{Ulysses}. Therefore, in this study, we used tomographic interplanetary scintillation (IPS) measurements, which can derive a solar-wind velocity map over all latitude ranges (\citeauthor{Kojima1998JGR}, \citeyear{Kojima1998JGR}). The data and the analysis method will be explained in Section \ref{Data}.

\section{Data and Analysis }  \label{Data}
\begin{figure}    
\centerline{\includegraphics[width=0.8\textwidth,clip=]{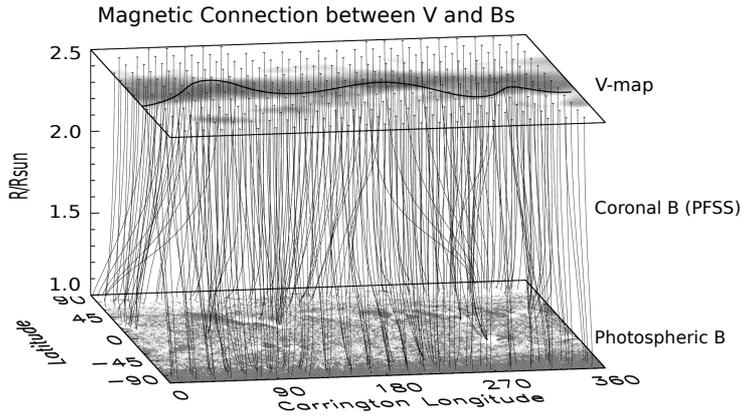}}\caption{A V-map on the source surface at 2.5 $\mathrm{R}_\odot$ is derived by the IPS-tomography method. Potential magnetic-field lines connect the V-map and the photosphere to obtain the flux expansion factor and $B_{\mathrm{S}}$.} \label{Fig001}
\end{figure}

In this study we used three data sets of  $V$,  $B_{\mathrm{S}}$ and $f$. We used the solar-wind velocities measured at Solar-Terrestrial Environment Laboratory (STEL) using the IPS method because the IPS measurements can observe the solar wind at all latitude ranges in a relatively short time, and the observations cover several solar cycles (\citeauthor{Kojima1990SSRv}, \citeyear{Kojima1990SSRv}). However, the solar-wind velocities that are estimated directly from the IPS measurement involve the line-of-sight integration effect. Therefore, we used the tomographic analysis technique (\citeauthor{Kojima1998JGR}, \citeyear{Kojima1998JGR}), which can deconvolve the line-of-sight integration effects and retrieve the intrinsic solar-wind velocity. Using this technique, we derived velocity distribution maps of latitude and longitude (V-map) on the source surface at 2.5 $\mathrm{R}_\odot$ (Figure \ref{Fig001}). Since IPS tomographic analysis requires good coverage of line-of-sight projection on the source surface, we used IPS data obtained during three Carrington rotations to derive one V-map. Therefore, each V-map represents the average structure during three Carrington rotations. This data-window of three rotations was shifted by one rotation to derive the map of the next rotation. Thus, we obtained about ten maps each year except for the snowy winter seasons when the IPS observations at STEL were interrupted. The analyses were made for the years from 1986 to 2009 over two solar cycles.  

In order to obtain magnetic-field properties of $B_{\mathrm{S}}$ and $f$, we used the solar magnetic field data observed at the National Solar Observatory/Kitt Peak, and calculated the coronal magnetic field configuration by the potential-field-source-surface (PFSS) model (\citeauthor{Schatten1969SoPh}, \citeyear{Schatten1969SoPh}; \citeauthor{Altschuler1969SoPh}, \citeyear{Altschuler1969SoPh}). The PFSS analysis code was developed by \citeauthor{Hakamada1999SoPh} (\citeyear{Hakamada1999SoPh}) and was used for mapping a solar-wind structure obtained by IPS to its counterpart on the photosphere (\citeauthor{Ohmi2001JGR}, \citeyear{Ohmi2001JGR}; \citeauthor{Kojima2004ASSL}, \citeyear{Kojima2004ASSL}). We traced back the velocity map from the source surface to the photosphere along the potential-field line to find the source region of the flow on the photosphere and obtain the photospheric magnetic-field strength $B_{\mathrm{S}}$ (Figure \ref{Fig001}). The flux expansion factor [$f$] is defined by the ratio between the magnetic-field strength on the photosphere and that on the source surface normalized by $(R_\mathrm{SS}/\mathrm{R}_\odot)^2$ , where  $R_\mathrm{SS}$ is the radius of the source surface. The flux expansion factors and the photospheric magnetic-field strength were averaged from  five solar-wind velocity ranges of 300\,--\,400, 400\,--\,500, 500\,--\,600, 600\,--\,700, and $>$700 $\mathrm{km\,s^{-1}}$. Thus, the averages of $f(V_\mathrm{i})$, and $B_{\mathrm{S}}(V_\mathrm{i})$ obtained for each Carrington rotation were averaged again over a year, from which we derived the year-averaged parameters $f$, $B_{\mathrm{S}}$, and the ratio between these two parameters [$B_{\mathrm{S}}/f$] as functions of $V$.

\section{Flux Expansion Factor [$f$]} \label{flux}
\begin{figure}    
\centerline{\includegraphics[width=0.9\textwidth]{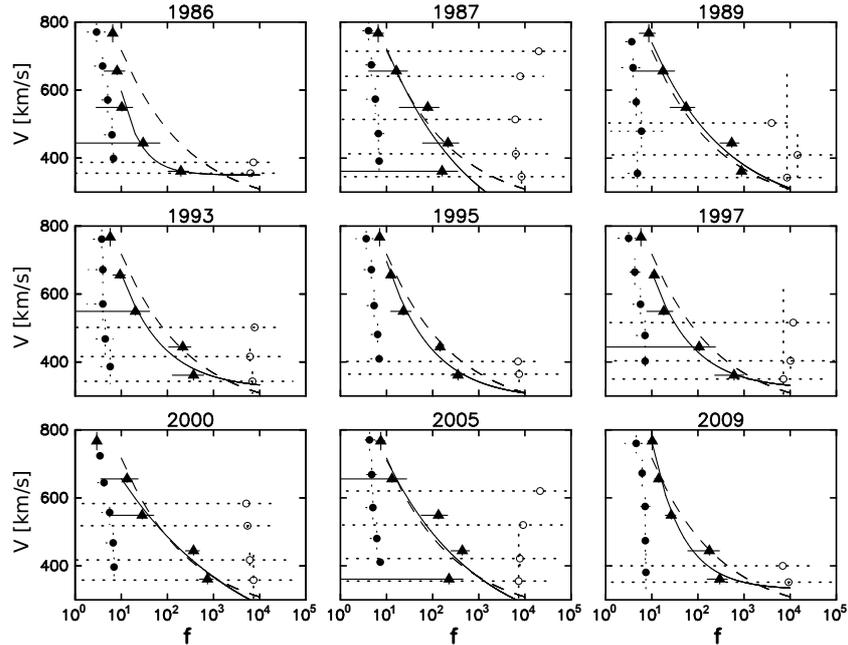}}
 \caption{Dependence of velocity on the flux-expansion factor. Triangles are averages of all data in each year. Solid circles are data with weak magnetic-field strength ($<$5 [G]) and small expansion factors ($<$100), while open circles are those with strong magnetic-field strength ($>$50 [G]) and a large expansion factor ($>$1000). A solid curved line is the best-fit solution for Equation (\ref{eq02}), while a dashed line is the average best-fit solution for the 24 years of data from year 1986 to 2009. } \label{Fig002}
\end{figure}

Figure \ref{Fig002} shows the dependence of velocity on the flux-expansion factor for the years 1986, 1987, 1989, 1993, 1995, 1997, 2000, 2005, and 2009. Triangles with solid error bars represent averages of all data in each year. We chose as examples the data of nine years among the 24 years, including the solar minimum and solar maximum phases. Solid curved lines represent the best-fit solution for  

\begin{equation} \label{eq02}
V=V_0+\alpha/f^\beta  ,                
\end{equation}
while the dashed lines represent the best-fit solution averaged for all data during 24 years given by 
\begin{equation} \label{eq03}
V=250+932/f^{0.30}                  
\end{equation}

The inverse relation between velocity and the flux-expansion factor was originally found by \citeauthor{Wang1990ApJ} (\citeyear{Wang1990ApJ,Wang1991ApJ}), and its inverse equation has been improved by \citeauthor{Arge2000JGR} (\citeyear{Arge2000JGR}) for solar-wind prediction. The original Wang and Sheeley equation was based upon the empirical correlation between the velocities observed at the Earth and the flux-expansion factor derived with the data measured at the Wilcox Solar Observatory. On the other hand, \citeauthor{Arge2000JGR} (\citeyear{Arge2000JGR}) propagated the solar-wind predicted on the source surface using Equation (\ref {eq02}) to the Earth, and then made iterations changing parameters in the equation until a reasonably good agreement was obtained between \insitu observations and predicted velocities at the Earth. Although Equation (\ref{eq03}) can be applied for all latitudes, the WSA equation should be applied for the observations near the solar Equator because it was derived by making an iterative comparison with the low-latitude \insitu data. The numerators of the second term in Equation (\ref{eq03}) and the WSA equation exibit a two-fold difference. This was caused mainly by the difference in the magnetic-field data, and therefore Equation (\ref{eq03}) should be applied to Kitt Peak data, which were measured with a higher resolution than those of the Wilcox Solar Observatory.

\begin{figure}    
\centerline{\includegraphics[width=0.8\textwidth]{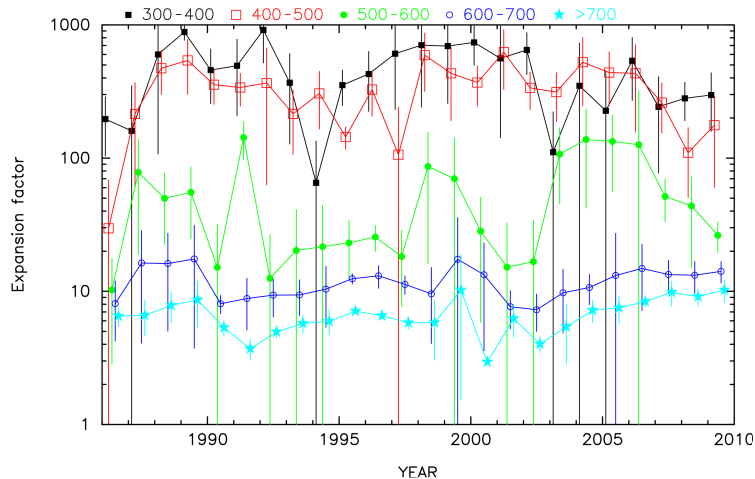}}
 \caption{Yearly variation in the expansion factors for the five solar-wind velocity ranges of 300\,--\,400, 400\,--\,500, 500\,--\,600, 600\,--\,700, and $>$ 700 \kms, respectively.}
  \label{Fig003}
\end{figure}

Figure \ref{Fig003} shows yearly variations in the expansion factors derived for the five solar-wind velocity ranges of 300\,--\,400, 400\,--\,500, 500\,--\,600, 600\,--\,700, and $>$ 700 \kms. We compared the flux expansions between the last solar minimum (2007\,--\,2009) and the previous minimum (1994\,--\,1996) in Table \ref{table 1}. The differences were  less than one-$\sigma$ variances of the all-year averages, but between two minimum phases, there were small but statistically significant differences in medium- and high-velocity winds. In the last solar minimum, $f$ increased 1.1 to 1.4 times, and this increase was similar to that reported by \citeauthor{Wang2013ApJ} (\citeyear{Wang2013ApJ}), who compared \textit{Ulysses} polar passes in 1993\,--\,1996 and 2006\,--\,2008, finding an increase of 1.2 times and a decrease in the median wind velocity by 3\,\% (from  740 \kms).

\begin{table}    
\caption{Average expansion factors for the five solar-wind  velocity ranges of 300\,--\,400, 400\,--\,500, 500\,--\,600, 600\,--\,700, and $>$ 700 \kms, respectively. The second row is the average through all years [$f_\mathrm{all}$], and the third and fourth rows are averages for three-year periods: 1994\,--\,1996 [$f_{1}$] and 2007\,--\,2009 [$f_{2}$]. The last row shows the ratio between two periods.}  \label{table 1}
\begin{tabular}{lccccc}
\hline
$V$ [\kms]      & 300\,--\,400        & 400\,--\,500        & 500\,--\,600        & 600\,--\,700        & $>$ 700 \\ \hline
$f_\mathrm{all}$    & 373$\pm$242   & 250$\pm$151   & 31$\pm$28     & 11$\pm$3      & 6$\pm$2 \\
$f_1$ & 221$\pm$159   & 195$\pm$79    & 24$\pm$2      & 12$\pm$1      & 7$\pm$0.4 \\
$f_2$ & 276$\pm$21    & 167$\pm$63    & 35$\pm$11     & 14$\pm$0.4    & 10$\pm$0.5 \\
$f_2/f_1$ & 1.25$\pm$0.58 & 0.86$\pm$0.65 & 1.46$\pm$0.22 & 1.17$\pm$0.08 & 1.43$\pm$0.05 \\
\hline
\end{tabular}
\end{table}

In order to compare this $V$--$f$ correlation with two other correlations of $V$--$B_{\mathrm{S}}$ and $V$--$B_{\mathrm{S}}/f$ in the following sections, we selected two kinds of solar wind. One originates from the region with weak magnetic-field strength ($<$5 [G]) and a small expansion factor ($<$100) (solid circles with dotted error bars in Figure \ref{Fig002}, hereafter referred to SW1), and the other one from the region with strong magnetic-field strength ($>$50 [G]) and a large expansion factor ($>$1000) (open circles with dotted error bars, hereafter referred to as SW2). We note in Figure \ref{Fig002} that these data not only largely deviate from the average group, but are also less dependent on $f$.

\section{Photospheric Magnetic Field Strength [$B_\mathrm{S}$]}   \label{Bphoto}
\begin{figure}    
   \centerline{\includegraphics[width=0.9\textwidth]{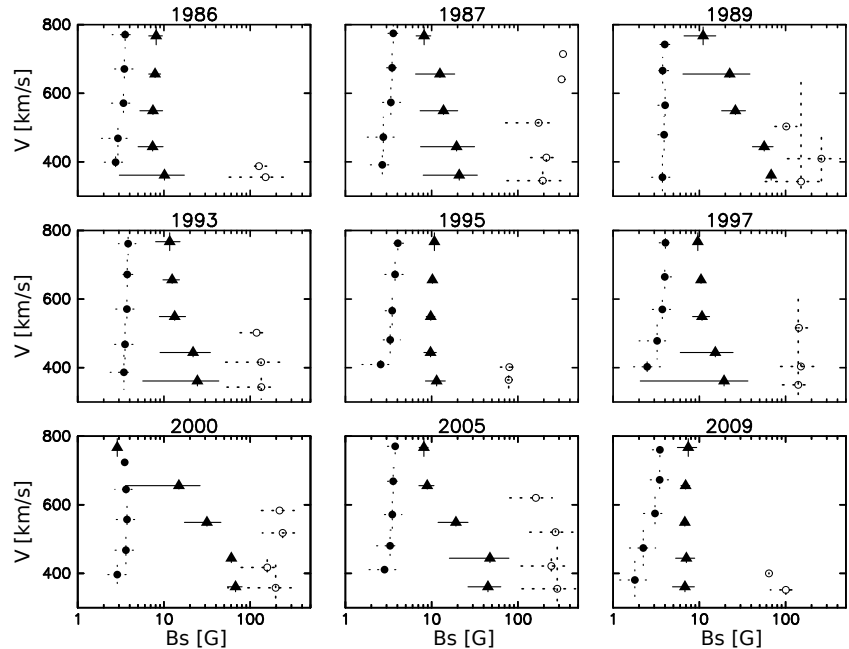}}
 \caption{Dependence of velocity on the photospheric magnetic-field strength. Triangles are averages of all data in each year. Solid circles are data with weak magnetic-field strength ($<$5 [G]) and a small expansion factor ($<$100), while open circles are data with strong magnetic-field strength ($>$50 [G]) and a large expansion factor ($>$1000). }
  \label{Fig004}
\end{figure}

\begin{figure}    
   \centerline{\includegraphics[width=0.8\textwidth]{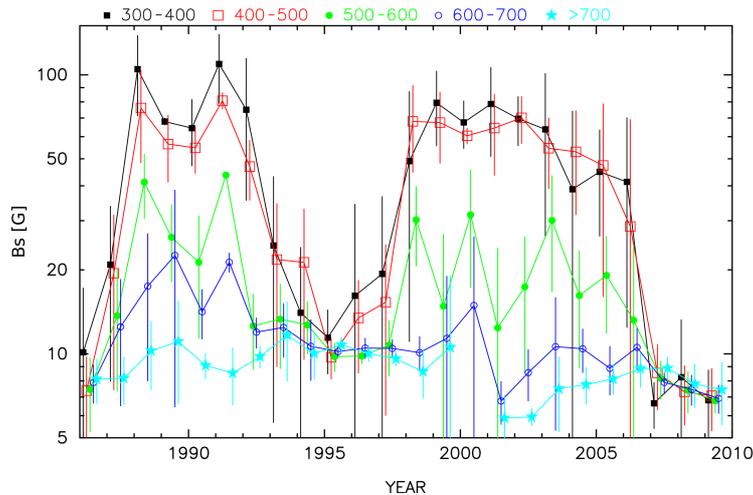}}
  \caption{Yearly variation of the magnetic-field strength of photospheric source regions for the five solar-wind velocity ranges same as Figure \ref{Fig003}. }
  \label{Fig005}
\end{figure}

Figure \ref{Fig004} shows examples of the dependence of velocity on the photospheric magnetic-field strength for the years 1986, 1987, 1989, 1993, 1995, 1997, 2000, 2005, and 2009. Triangles with solid error bars are averages of all data in each year. An inverse relation between $V$ and $B_{\mathrm{S}}$ exists, but not obviously in the three solar minima (1986, 1995, and 2009). Figure \ref{Fig005} shows yearly variation in the magnetic-field strength derived for the five solar-wind velocity ranges of 300\,--\,400, 400\,--\,500, 500\,--\,600, 600\,--\,700, and  $>$ 700 \kms. The average $B_{\mathrm{S}}$ during the three years around the two solar minimum phases of 1994 to 1996 and 2007 to 2009 are shown in Table \ref{table B}. In the solar-minimum phase, $B_{\mathrm{S}}$ has a very small range of 10.3\,--\,12.5 [G] and 6.9\,--\,8.2 [G]. This is one of the reasons why the inverse relations between $V$ and $B_{\mathrm{S}}$ (triangle) are not obvious in the three solar minima. Comparing the two minimum phases, the field strength of the last minimum phase decreased about 30\,\% (Table \ref{table B}) of that of the previous minimum. \citeauthor{Wang2013ApJ} (\citeyear{Wang2013ApJ}) also reported a 31\,\% decrease from 10.6 to 7.3 [G] in footpoint field strength of the polar wind.

\begin{table}    
\caption{Average photospheric magnetic-field strength for the five solar-wind velocity ranges same as Table \ref{table 1} during three-year periods: 1994\,--\,1996 [$B_1$] and 2007\,--\,2009 [$B_2$]. The last row shows the ratio of the average magnetic-field strength between two periods.}  \label{table B}
\begin{tabular}{lccccc}
\hline
$V$ [\kms]      & 300\,--\,400        & 400\,--\,500        & 500\,--\,600        & 600\,--\,700        & $>$ 700 \\ \hline
$B_1$ [G] & 12.5$\pm$1.7 & 11.6$\pm$3.5 & 10.2$\pm$1.0 & 10.3$\pm$0.2 & 10.3$\pm$0.4 \\
$B_2$ [G] & 6.9$\pm$0.5  & 7.6$\pm$0.6  & 7.3$\pm$0.6  & 7.5$\pm$0.4  & 8.2$\pm$0.6 \\
$B_2/B_1$ & 0.6$\pm$0.3  & 0.7$\pm$0.5  & 0.7$\pm$0.2  & 0.7$\pm$0.1  & 0.8$\pm$0.1 \\
\hline
\end{tabular}
\end{table}

\begin{figure}    
\centerline{\includegraphics[width=0.9\textwidth,clip=]{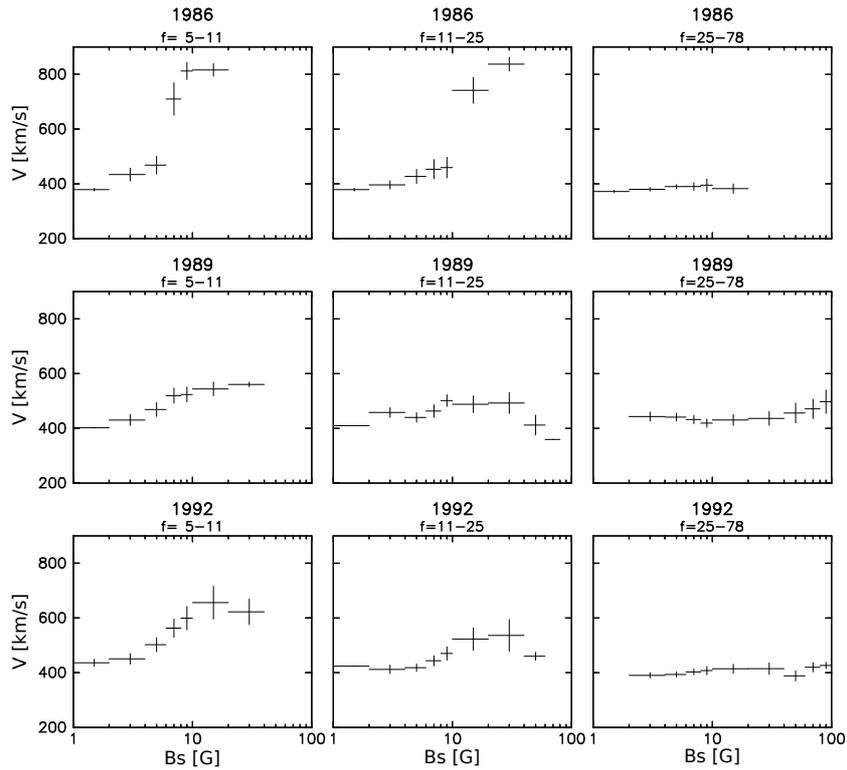}}
  \caption{Dependence of velocity on the photospheric magnetic-field strength. Data are classified into three expansion-factor bins, 5\,--\,11, 11\,--\,25, and 25\,--\,78. The three years of 1986, 1989, and 1992 are illustrated.}
  \label{Fig006}
\end{figure}

The FSK equation indicates that the solar-wind velocity at the Earth should have positive dependence on the photospheric-field strength. However, we cannot see this tendency in Figure \ref{Fig004}, and furthermore \citeauthor{Wang1996Sci} (\citeyear{Wang1996Sci}), \citeauthor{Wang2009ApJ} (\citeyear{Wang2009ApJ}), and \citeauthor{Wang2013ApJ} (\citeyear{Wang2013ApJ}) reported that there appears to be almost no systematic relation between them. However, the SZK equation indicates that the additional acceleration term has a positive dependence on $B_{\mathrm{S}}$. Since this term includes not only $B_{\mathrm{S}}$ but also some other physical parameters such as $f$ and the mass flux, we examined the dependence on $B_{\mathrm{S}}$, classifying the data by the expansion factor [$f$]. Data are classified into three expansion factors of 5\,--\,11, 11\,--\,25, and 25\,--\,78. Figure \ref{Fig006} shows as examples three years: 1986, 1989, and 1992, which include solar minimum and maximum phases. The figures show that velocity increases with $B_{\mathrm{S}}$ when the flux expansion is small, and the dependency becomes less obvious as the flux expansion increases. This behavior can be understood by the first term in Equation (\ref{eqSZK}); as $f$ becomes large, the first term, which includes $B_{\mathrm{S}}/f$, becomes less effective for determining the final velocity.

In Figure \ref{Fig004} we showed two special kinds of solar wind (solid and open circles). Solid circles are the SW1 with weak magnetic-field strength ($<$5 [G]) and a small expansion factor ($<$100), and open circles are SW2 with strong magnetic-field strength ($>$50 [G]) and a large expansion factor ($>$1000). We note that SW1 (solid circle) have a positive relation between $V$ and $B_{\mathrm{S}}$, unlike the data of the total average (triangles). 

\section{Ratio of Magnetic Field Strength and Flux Expansion Factor [$B_{\mathrm{S}}/f$]} \label{Bf}
\begin{figure}    
   \centerline{\includegraphics[width=1.0\textwidth]{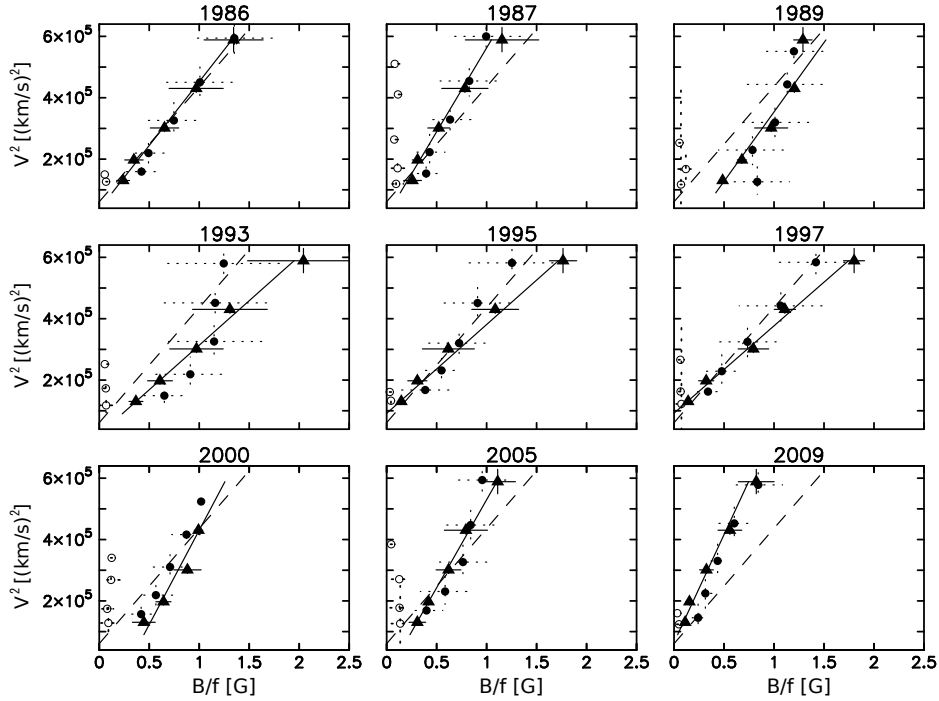}}
   \caption{Dependence of velocity on the ratio of photospheric magnetic-field strength and a flux-expansion factor. Triangles are averages of all data in a year. Solid circles are SW1 with weak magnetic-field strength ($<$5 [G]) and a small expansion factor ($<$100), and open circles are SW2 with strong magnetic-field strength ($>$50 [G]) and a large expansion factor ($>$1000). Solid lines are the best-fit regression lines, while dashed lines are the average for the 24 years of data from 1986 to 2009. }
  \label{Fig007}
\end{figure}

Since the velocity is dependent not only on the inverse of the flux expansion factor [$1/f$] but also on the photospheric magnetic-field strength [$B_{\mathrm{S}}$], we here combined these two parameters as $B_{\mathrm{S}}/f$ to examine the relation with velocity. Practically the parameter $B_{\mathrm{S}}/f$ can be replaced by the magnetic field at the source surface [$B_\mathrm{SS}$] in the PFSS analysis.
Figure \ref{Fig007} shows as examples the dependence of velocity on $B_{\mathrm{S}}/f$ for the years 1986, 1987, 1989, 1993, 1995, 1997, 2000, 2005, and 2009. According to the SZK equation, the correlations are shown for $V^2$. Solid lines represent the best-fit line for the equation of 
\begin{equation} \label{eq04}
 V^2 = V_0^2 + \alpha \frac{B_{\mathrm{S}}}{f} ,                
 \end{equation}
while dashed ones represent
\begin{equation} \label{eq05}
 V^2 = 247^2 + 613^2 \frac{B_{\mathrm{S}}}{f}  ,             
 \end{equation}
which was obtained by averaging 24 years of data. Although the best-fit line changes year by year, $V^2$ had a fairly good linear dependence on $B_{\mathrm{S}}/f$ throughout the 24 years. 

In order to see the features of the correlation $V$-$B_{\mathrm{S}}/f$ compared with the other two correlations of $V$-$f$ and $V$-$B_{\mathrm{S}}$, we note the two special kinds of solar wind of SW1 and SW2 shown by open and closed circles in Figure \ref{Fig007}. Although these kinds of solar wind largely deviated from the major average groups in $f$ (Figure \ref{Fig002}) and $B_{\mathrm{S}}$ (Figure \ref{Bphoto}) correlation diagrams, the SW1 data fit well on the linear relation with $B_{\mathrm{S}}/f$. However, SW2 does not have an obvious dependence on $B_{\mathrm{S}}/f$, having a small value of $B_{\mathrm{S}}/f$.  

\begin{figure}    
   \centerline{\includegraphics[width=0.8\textwidth]{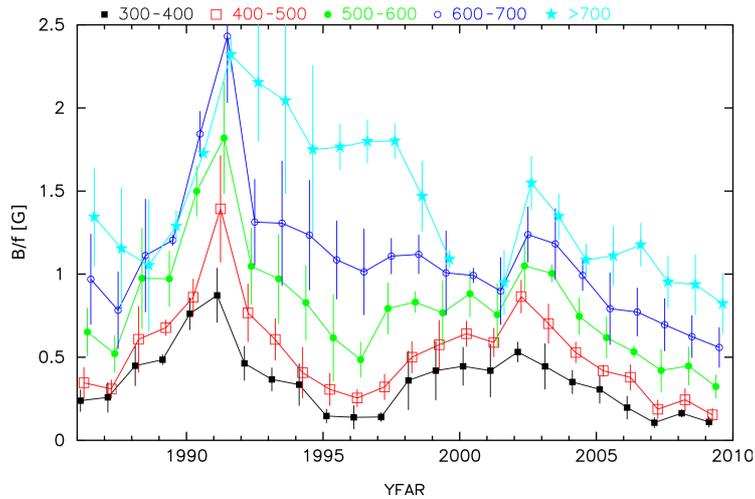}}
\caption{Yearly variation of $B_{\mathrm{S}}/f$ for the five solar-wind velocity ranges same as Figure \ref{Fig003}.}
  \label{Fig008}
\end{figure}

Figure \ref{Fig008} shows the yearly variation of $B_{\mathrm{S}}/f$ derived for the five solar-wind velocity ranges of 300\,--\,400, 400\,--\,500, 500\,--\,600, 600\,--\,700, and  $>$ 700 \kms, and Table \ref{table bf} compares variations between the two solar minimum phases. $B_{\mathrm{S}}/f$ in the last minimum phase decreased from that in the previous phase, and high-velocity winds (600\,--\,700 and $>$ 700 \kms)  especially decreased by as much as 50\,\%. The major percent of this decrease was caused by the 30\,\% decrease of $B_{\mathrm{S}}$ (Table \ref{table B}) and the 10 to 40\,\% increase of $f$ (Table \ref{table 1}). 

\begin{table}    
\caption{Average $B_{\mathrm{S}}/f$ for the five solar-wind velocity ranges same as Table \ref{table 1} during three-year periods: 1994\,--\,1996 [$(B_\mathrm{S}/f)_1$] and 2007\,--\,2009 [$(B_\mathrm{S}/f)_2$]. The last row shows the ratio between two periods.}  \label{table bf}
\begin{tabular}{lccccc}
\hline
$V$ [\kms]      & 300\,--\,400        & 400\,--\,500        & 500\,--\,600        & 600\,--\,700        & $>$ 700 \\ \hline
$(B_{\mathrm{S}}/f)_1$ & 0.176$\pm$0.072 & 0.297$\pm$0.056 & 0.599$\pm$0.142 & 1.101$\pm$0.087 & 1.778$\pm$0.019 \\
$(B_{\mathrm{S}}/f)_2$ & 0.131$\pm$0.027 & 0.187$\pm$0.036 & 0.383$\pm$0.056 & 0.620$\pm$0.055 & 0.906$\pm$0.058 \\
$(B_{\mathrm{S}}/f)_2/(B_{\mathrm{S}}/f)_1$  & 0.74$\pm$0.62   & 0.63$\pm$0.43   & 0.64$\pm$0.44   & 0.56$\pm$0.21   & 0.51$\pm$0.13 \\
\hline
\end{tabular}
\end{table}

\section{Discussion}  \label{discussion}
We investigated the relation between the solar-wind velocity and the coronal magnetic parameters from 1986 to 2009. From the cross-correlation analysis, we verified that $V$ has an inverse relation with $f$ (Equation (\ref{eq03})) as reported by \citeauthor{Wang1990ApJ} (\citeyear{Wang1990ApJ,Wang1991ApJ}). There are reports that both the mass and energy-flux densities tend to increase approximately linearly with increasing magnetic-field strength, but there appears to be almost no systematic relation between photospheric magnetic-field strength and the solar-wind velocity at the Earth (\citeauthor{Wang1996Sci}, \citeyear{Wang1996Sci}; \citeauthor{Wang2009ApJ} \citeyear{Wang2009ApJ}; \citeauthor{Wang2013ApJ}, \citeyear{Wang2013ApJ}). However, we found a different relation between photospheric magnetic-field strength and the solar-wind velocity. Investigating the $V$--$B_{\mathrm{S}}$ relations for the solar wind binned by the flux expansion factor, we found that the higher velocity winds flow out of stronger field regions in each bin. Combining these two relations of $V$--$f$ and $V$--$B_{\mathrm{S}}$, we revealed that the $B_{\mathrm{S}}/f$ has an extremely good linear correlation with the square of the velocity (Equation \ref{eq05}). This linear correlation also holds for the solar wind from CHs that have weak magnetic field and a small expansion factor, although they substantially deviated from the $V$--$f$ correlation. 

\begin{figure}    
\centerline{\includegraphics[width=0.9\textwidth]{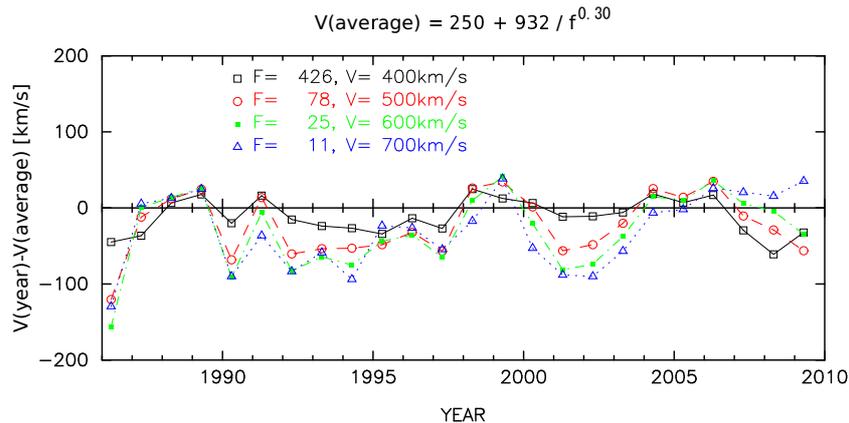}}
\caption{Differences in the best-fit solutions between the solid line and the dotted line in Figure \ref{Fig002}. The discrepancies are shown for four different expansion factors of 2663, 489, 160, and 69, which give velocities of 400, 500, 600, and 700 \kms from Equation (\ref{eq03}), respectively.}  \label{Fig009}
\end{figure}

We compared Equations (\ref{eq03}) and (\ref{eq05}), which are better models for making solar-wind predictions through the whole solar cycle. Since these equations were derived from averaging 24 years of data, we examined discrepancies in them from the equations obtained for each year (discrepancies between the dashed line and the solid line in Figures \ref{Fig002} and \ref{Fig007}). Figure \ref{Fig009} examines Equation (\ref{eq03}). The discrepancies are shown for four different expansion factors of 426, 78, 25, and 11 which give velocities of 400, 500, 600, and 700 \kms from Equation (\ref{eq03}), respectively. The average discrepancies over the 24 years are $-11\pm23$, $-22\pm40$, $-30\pm48$, and $-30\pm48$ \kms at each of the four expansion factors. This means that a prediction over the whole solar cycle can be made accurately using the single equation. In particular, the prediction of low- and medium-velocity solar winds, which are usually observed at the solar Equator, can be made with accuracy  better than 50 \kms.

\begin{figure}    
   \centerline{\includegraphics[width=0.9\textwidth]{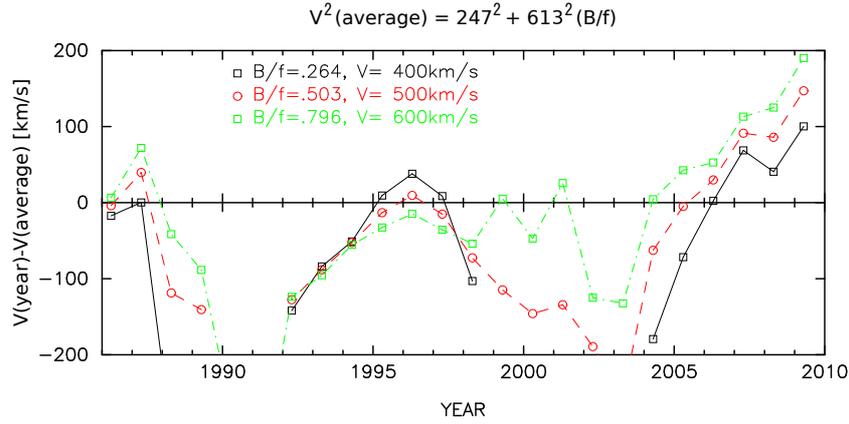}}
\caption{Difference in the best-fit solutions between the solid line and the dotted line in Figure \ref{Fig007}. The discrepancies are shown for three values of $B_{\mathrm{S}}/f$ of 0.068, 0.102, and 0.144, which give velocities of 400, 500, and 600 from Equation (\ref{eq05}), respectively.}
  \label{Fig010}
\end{figure}

Figure \ref{Fig010} examines Equation (\ref{eq05}). Discrepancies between dashed lines and solid lines in Figure \ref{Fig007} are shown for three different values of $B_{\mathrm{S}}/f$: 0.264, 0.503, and 0.796, which give velocities of 400, 500, and 600 \kms \ from Equation (\ref{eq05}). Figure \ref{Fig010} indicates that this single equation cannot be applied for solar-wind prediction through the whole solar cycle. This is because not only $B_{\mathrm{S}}/f$ but also the parameter $\alpha$ change significantly from year to year. 
This result reflects following two facts: i) the parameter $B_{\mathrm{S}}/f$ which is exactly same as the magnetic field strength on the source surface in this analysis determines the global magnetic structure in interplanetary space and depends mainly on solar magnetic dipole moment which changes significantly through solar cycle, and ii) the range of solar wind velocity hardly changes through solar cycle.

\begin{figure}    
   \centerline{\includegraphics[width=0.9\textwidth]{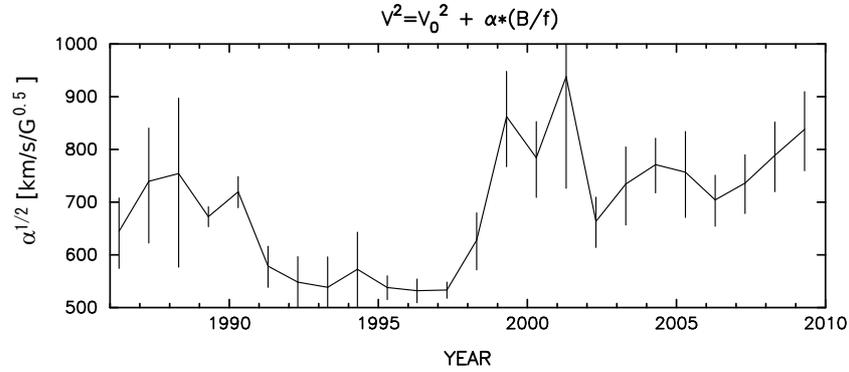}}
              \caption{Yearly variation of the gradient [$\alpha$] of Equation (\ref{eq04}).}
  \label{Fig011}
\end{figure}

It has been reported that the polar solar wind decreased in velocity in 2006\,--\,2008 compared with that observed in 1993\,--\,1996, but the decrease was as small as 3\,\% (from 763 to 740 \kms) (\citeauthor{McComas2008GeoRL}, \citeyear{McComas2008GeoRL}; \citeauthor{Wang2013ApJ}, \citeyear{Wang2013ApJ}). On the other hand, the changes of $B_{\mathrm{S}}$, $f$ and $B_{\mathrm{S}}/f$ between the two minimum phases were not small. In particular, the change of the parameter $\alpha$ in Equation (\ref{eq04}) was as large as 200\,\% between the two minimum phases (Figure \ref{Fig011} and Table \ref{table 3}). The less significant change of the polar high-velocity wind between two minimum phases can be understood by the second term of Equation (\ref{eq04}), $\alpha B_{\mathrm{S}}/f$, which gives additional acceleration to the thermal-driven solar wind. Since $\alpha$ increased 200\,\% and $B_{\mathrm{S}}/f$ decreased about 50\,\% (Table \ref{table bf}), the change of the additional acceleration is small. 

\begin{table}    
\caption{Comparison of averaged parameters $V_0$ and $\alpha$ in Equation (\ref{eq04}) between two solar minimum phases, The third column represents avreaged $B_{\mathrm{S}}/f$ for $V=750$ \kms. The last row shows the ratio between the two periods.}  \label{table 3}
\begin{tabular}{lcccc}
\hline
period    & $V_0$ & $\alpha$ & $B_{\mathrm{S}}/f$ [750 \kms]  \\ \hline
1994\,--\,1996 & 321$\pm$20    & $535^2$$\pm$$57^2$  & 1.778$\pm$0.019  \\
2007\,--\,2009 & 237$\pm$43    & $782^2$$\pm$$255^2$ & 0.906$\pm$0.058  \\
$\frac{2007-2009}{1994-1996}$ & 0.74$\pm$0.26 & 2.13$\pm$0.05 & 0.51$\pm$0.13  \\
\hline
\end{tabular}
\end{table}

The square of the first term in Equation (\ref{eq05}), $V_0^2$, corresponds to $ 2 [\gamma/(\gamma-1) \mathrm{R} T_\mathrm{c} - (\mathrm{G} \mathrm{M}_\odot/\mathrm{R}_\odot)]$ in the SZK equation. Therefore, $V_0^2 /2 + \mathrm{G} \mathrm{M}_\odot/R_\odot$ is proportional to the coronal temperature [$T_\mathrm{c}$], and the $V_0$ decrease of 25\,\% between two minimum phases (Table \ref{table 3}) corresponds to a 6\,\% decrease of coronal temperature. This is the same freeze-in temperature decrease derived from the ion charge ratio of solar wind by \citeauthor{vonSteiger2011JGRA} (\citeyear{vonSteiger2011JGRA}).

According to the SZK equation, $\alpha$ varies in proportion to the inverse of mass flux. Although the solar-wind velocity did not change significantly between the two minimum phases, solar-wind densities provided by the NASA/GSFC's Space Physics Data Facility's OMNIweb service decreased from 30 to 40\,\% (Table \ref{table n}). \citeauthor{McComas2008GeoRL} (\citeyear{McComas2008GeoRL}) also reported that the fast solar wind from both large polar CHs was significantly less dense (17\,\%) than during the previous solar minimum. Therefore, half of the $\alpha$ increase could be attributed to this density decrease. Since $\alpha$ also consists of the Poynting flux from the photosphere [$\langle \delta B_{\perp} \delta v_{\perp} \rangle_\mathrm{S}$], another half of the $\alpha$ increase may have been caused from an increase of the Poynting flux. From this viewpoint, it is worth noting that \citeauthor{Wang1991ApJ} (\citeyear{Wang1991ApJ}) commented that if Alfv\'en waves were responsible for accelerating the solar wind, and the wave energy flux is roughly constant, the mean-square wave velocity [$\langle \delta v_0^2 \rangle$] and the field strength [$B_{\mathrm{S}}$] at the coronal base may be inversely correlated.

\begin{table}    
\caption{Average solar-wind density at 1 AU for the five solar-wind velocity ranges same as Table \ref{table 1} during three-year periods: 1994\,--\,1996 [$N_1$] and 2007\,--\,2009 [$N_2$]. The last row shows the ratio between the two periods.}  \label{table n}
\begin{tabular}{lccccc}
\hline
$V$ [\kms]      & 300\,--\,400        & 400\,--\,500        & 500\,--\,600        & 600\,--\,700        & $>$ 700 \\ \hline
$N_1$ [$\mathrm{cm}^{-3}$] & 11.13$\pm$6.29 & 6.70$\pm$4.30 & 4.29$\pm$2.17 & 3.49$\pm$1.34 & 3.19$\pm$2.32 \\
$N_2$ [$\mathrm{cm}^{-3}$] & 6.77$\pm$4.84  & 4.11$\pm$3.01 & 2.96$\pm$1.74 & 2.58$\pm$0.89 & 2.01$\pm$0.61 \\
$N_2/N_1$ & 0.61$\pm$3.85  & 0.61$\pm$2.68 & 0.69$\pm$1.55 & 0.74$\pm$1.02 & 0.63$\pm$1.47 \\
\hline
\end{tabular}
\end{table}

Making cross-correlation analyses between the solar-wind data measured by Advanced Composition Explorer (ACE) spacecraft and the photospheric magnetic-field strength, Wang and Sheeley (2013) obtained the following equation between the mass flux density at the coronal base and the footpoint field strength:

\begin{equation} \label{eq06}
n_0 v_0 \approx 8 \times 10^{12} \mathrm{cm^{-2} s^{-1}} \frac{B_{\mathrm{S}}}{1\, [\mathrm{G}]},         
\end{equation}
where $B_{\mathrm{S}}$ is a footpoint field strength. Therefore, if we compare the solar-wind streams which have the same flux expansion factor, $B_{\mathrm{S}}/n_\mathrm{e} v_\mathrm{e}$ is a constant value that does not depend on solar activity. In fact, Tables \ref{table B} and \ref{table n} show that the changes of $B_{\mathrm{S}}$ and $n_\mathrm{e} v_\mathrm{e}$ between the two minimum phases are the same amount of 30\,\%. Consequently, the mass flux and the magnetic-field strength disappear from the additional acceleration term in the SZK equation, becoming $\mathrm{const.} \times \langle \delta B_\perp \delta v_\perp \rangle_\mathrm{S} /f$, which is similar to the WSA equation except for the Poynting flux. This equation can be made equivalent to the WSA equation, by making the Poynting flux less dependent on the solar activity. However, we discussed above that the Poynting vector has to be about 1.4 times larger in the last minimum phase to make the additional acceleration term in the SZK equation less sensitive to solar activity. This contradiction must be investigated in a further study.

\section{Conclusions}  \label{conclusion}
We investigated experimentally the relation between the solar-wind velocity and the coronal magnetic parameters by comparing the WSA and SZK equations. We verified that $V$ is inversely proportional to $f$ and found that $V$ tends to increase with $B_{\mathrm{S}}$ if $f$ is the same. Consequently, we revealed that $V$ has an extremely good linear correlation with $B_{\mathrm{S}}/f$ as indicated by the SZK equation. However, we should note that the SZK equation ($V$--$B_{\mathrm{S}}/f$ relation) cannot be used for solar-wind prediction through the whole solar cycle. The slope of its linear relation changes year by year largely because it explicitly includes a mass flux. On the other hand, the WSA equation [$V$--$f$ relation] is a good prediction model, which can be used through the whole solar cycle, although it includes neither the mass flux nor $B_{\mathrm{S}}$ which play important physical roles for the additional acceleration of the solar wind. This is because the mass flux and magnetic-field strength can be canceled in the SZK equation using their linear relation.

The slope of the linear relation, $V$--$B_{\mathrm{S}}/f$, shows a solar-cycle variation. Especially we focused on the comparison between the two solar-minimum phases,  1994\,--\,1996 and 2007\,--\,2009. Consequently we found that the slope is two times larger in the very quiet solar minimum (2007\,--\,2009) in comparison with that in 1994\,--\,1996. 
This difference in solar-minimum phases is probably explained by a long-term variation of the solar-cycle activity.
However, the slope also tends to increase during solar active phases, which was not discussed in this article; we will extend this examination including the whole solar cycle in a further study.

\section*{Disclosure of Potential Conflicts of Interest}
The authors declare that they have no conflict of interest.

\begin{acks}
The IPS observations were carried out under the solar-wind program of the Solar-Terrestrial Environment Laboratory
(STEL) of Nagoya University, and were partially supported by the IUGONET Project of MEXT, Japan.
The authors would like to express their thanks for the use of the NSO/Kitt Peak magnetogram data. 
They are also indebted to the NASA/GSFC's Space Physics Data Facility's OMNIweb service, and OMNI data. 
This work was carried out by the joint research program of the STEL, Nagoya University, and was also partially supported by the JSPS Grant-in-Aid for Scientific Research A (25257079).
\end{acks}
 \bibliographystyle{spr-mp-sola}
 \bibliography{references}  

\begin{thebibliography}{23}
\ifx \bisbn   \undefined \def \bisbn  #1{ISBN #1}\fi
\ifx \binits  \undefined \def \binits#1{#1}\fi
\ifx \bauthor  \undefined \def \bauthor#1{#1}\fi
\ifx \batitle  \undefined \def \batitle#1{#1}\fi
\ifx \bjtitle  \undefined \def \bjtitle#1{\textit{#1}}\fi
\ifx \bvolume  \undefined \def \bvolume#1{\textbf{#1}}\fi
\ifx \byear  \undefined \def \byear#1{#1}\fi
\ifx \bissue  \undefined \def \bissue#1{#1}\fi
\ifx \bfpage  \undefined \def \bfpage#1{#1}\fi
\ifx \blpage  \undefined \def \blpage #1{#1}\fi
\ifx \burl  \undefined \def \burl#1{\textsf{#1}}\fi
\ifx \href  \undefined \def \href#1#2{\textsf{#2}}\fi
\ifx \doiurl  \undefined \def
  \doiurl#1{\href{http://dx.doi.org/#1}{\textsf{#1}}}\fi
\ifx \betal  \undefined \def \betal{\textit{et al.}}\fi
\ifx \binstitute  \undefined \def \binstitute#1{#1}\fi
\ifx \bctitle  \undefined \def \bctitle#1{#1}\fi
\ifx \beditor  \undefined \def \beditor#1{#1}\fi
\ifx \bpublisher  \undefined \def \bpublisher#1{#1}\fi
\ifx \bbtitle  \undefined \def \bbtitle#1{\textit{#1}}\fi
\ifx \bedition  \undefined \def \bedition#1{#1}\fi
\ifx \bseriesno  \undefined \def \bseriesno#1{\textbf{#1}}\fi
\ifx \blocation  \undefined \def \blocation#1{#1}\fi
\ifx \bsertitle  \undefined \def \bsertitle#1{\textit{#1}}\fi
\ifx \bsnm \undefined \def \bsnm#1{#1}\fi
\ifx \bsuffix \undefined \def \bsuffix#1{#1}\fi
\ifx \bparticle \undefined \def \bparticle#1{#1}\fi
\ifx \barticle \undefined \def \barticle#1{}\fi
\ifx \botherref \undefined \def \botherref#1{}\fi
\ifx \url \undefined \def \url#1{\textsf{#1}}\fi
\ifx \bchapter \undefined \def \bchapter#1{}\fi
\ifx \bbook \undefined \def \bbook#1{}\fi
\ifx \bcomment \undefined \def \bcomment#1{#1}\fi
\ifx \oauthor \undefined \def \oauthor#1{#1}\fi
\ifx \citeauthoryear \undefined \def \citeauthoryear#1{#1}\fi
\def \endbibitem {}
\ifx \bconflocation  \undefined \def \bconflocation#1{#1} \fi

\bibitem[\protect\citeauthoryear{{Altschuler} and
  {Newkirk}}{1969}]{Altschuler1969SoPh}
\begin{barticle}
\bauthor{\bsnm{{Altschuler}}, \binits{M.D.}},
\bauthor{\bsnm{{Newkirk}}, \binits{G.}}:
\byear{1969},
\batitle{{Magnetic Fields and the Structure of the Solar Corona. I: Methods of
  Calculating Coronal Fields}}.
\bjtitle{\solphys}
\bvolume{9},
\bfpage{131}\,--\,\blpage{149}.
doi:\doiurl{10.1007/BF00145734}.
\end{barticle}
\endbibitem

\bibitem[\protect\citeauthoryear{{Arge} and {Pizzo}}{2000}]{Arge2000JGR}
\begin{barticle}
\bauthor{\bsnm{{Arge}}, \binits{C.N.}},
\bauthor{\bsnm{{Pizzo}}, \binits{V.J.}}:
\byear{2000},
\batitle{{Improvement in the prediction of solar wind conditions using
  near-real time solar magnetic field updates}}.
\bjtitle{\jgr}
\bvolume{105},
\bfpage{10465}\,--\,\blpage{10480}.
doi:\doiurl{10.1029/1999JA000262}.
\end{barticle}
\endbibitem

\bibitem[\protect\citeauthoryear{{Fisk}}{2003}]{Fisk2003JGRA}
\begin{barticle}
\bauthor{\bsnm{{Fisk}}, \binits{L.A.}}:
\byear{2003},
\batitle{{Acceleration of the solar wind as a result of the reconnection of
  open magnetic flux with coronal loops}}.
\bjtitle{Journal of Geophysical Research (Space Physics)}
\bvolume{108},
\bfpage{1157}.
doi:\doiurl{10.1029/2002JA009284}.
\end{barticle}
\endbibitem

\bibitem[\protect\citeauthoryear{{Fisk}, {Schwadron}, and
  {Zurbuchen}}{1999}]{Fisk1999JGR}
\begin{barticle}
\bauthor{\bsnm{{Fisk}}, \binits{L.A.}},
\bauthor{\bsnm{{Schwadron}}, \binits{N.A.}},
\bauthor{\bsnm{{Zurbuchen}}, \binits{T.H.}}:
\byear{1999},
\batitle{{Acceleration of the fast solar wind by the emergence of new magnetic
  flux}}.
\bjtitle{\jgr}
\bvolume{104},
\bfpage{19765}\,--\,\blpage{19772}.
doi:\doiurl{10.1029/1999JA900256}.
\end{barticle}
\endbibitem

\bibitem[\protect\citeauthoryear{{Hakamada} and
  {Kojima}}{1999}]{Hakamada1999SoPh}
\begin{barticle}
\bauthor{\bsnm{{Hakamada}}, \binits{K.}},
\bauthor{\bsnm{{Kojima}}, \binits{M.}}:
\byear{1999},
\batitle{{Solar Wind Speed and Expansion Rate of the Coronal Magnetic Field
  during Carrington Rotation 1909}}.
\bjtitle{\solphys}
\bvolume{187},
\bfpage{115}\,--\,\blpage{122}.
doi:\doiurl{10.1023/A:1005183914772}.
\end{barticle}
\endbibitem

\bibitem[\protect\citeauthoryear{{Hirano}
  \textit{et~al.}}{2003}]{Hirano2003AGUFM}
\begin{botherref}
\oauthor{\bsnm{{Hirano}}, \binits{M.}},
\oauthor{\bsnm{{Kojima}}, \binits{M.}},
\oauthor{\bsnm{{Tokumaru}}, \binits{M.}},
\oauthor{\bsnm{{Fujiki}}, \binits{K.}},
\oauthor{\bsnm{{Ohmi}}, \binits{T.}},
\oauthor{\bsnm{{Yamashita}}, \binits{M.}},
\oauthor{\bsnm{{Hakamada}}, \binits{K.}},
\oauthor{\bsnm{{Hayashi}}, \binits{K.}}:
2003,
{The Relation Between the Solar Wind Velocity and the Magnetic Conditions of
  Coronal Holes}.
\textit{AGU Fall Meeting Abstracts},
B164.
\end{botherref}
\endbibitem

\bibitem[\protect\citeauthoryear{{Kojima} and
  {Kakinuma}}{1990}]{Kojima1990SSRv}
\begin{barticle}
\bauthor{\bsnm{{Kojima}}, \binits{M.}},
\bauthor{\bsnm{{Kakinuma}}, \binits{T.}}:
\byear{1990},
\batitle{{Solar cycle dependence of global distribution of solar wind speed}}.
\bjtitle{\ssr}
\bvolume{53},
\bfpage{173}\,--\,\blpage{222}.
doi:\doiurl{10.1007/BF00212754}.
\end{barticle}
\endbibitem

\bibitem[\protect\citeauthoryear{{Kojima}
  \textit{et~al.}}{1998}]{Kojima1998JGR}
\begin{barticle}
\bauthor{\bsnm{{Kojima}}, \binits{M.}},
\bauthor{\bsnm{{Tokumaru}}, \binits{M.}},
\bauthor{\bsnm{{Watanabe}}, \binits{H.}},
\bauthor{\bsnm{{Yokobe}}, \binits{A.}},
\bauthor{\bsnm{{Asai}}, \binits{K.}},
\bauthor{\bsnm{{Jackson}}, \binits{B.V.}},
\bauthor{\bsnm{{Hick}}, \binits{P.L.}}:
\byear{1998},
\batitle{{Heliospheric tomography using interplanetary scintillation
  observations 2. Latitude and heliocentric distance dependence of solar wind
  structure at 0.1-1 AU}}.
\bjtitle{\jgr}
\bvolume{103},
\bfpage{1981}\,--\,\blpage{1990}.
doi:\doiurl{10.1029/97JA02162}.
\end{barticle}
\endbibitem

\bibitem[\protect\citeauthoryear{{Kojima}
  \textit{et~al.}}{2004}]{Kojima2004ASSL}
\begin{bchapter}
\bauthor{\bsnm{{Kojima}}, \binits{M.}},
\bauthor{\bsnm{{Fujiki}}, \binits{K.}},
\bauthor{\bsnm{{Hirano}}, \binits{M.}},
\bauthor{\bsnm{{Tokumaru}}, \binits{M.}},
\bauthor{\bsnm{{Ohmi}}, \binits{T.}},
\bauthor{\bsnm{{Hakamada}}, \binits{K.}}:
\byear{2004},
\bctitle{{Solar Wind Properties from IPS Observations}}.
In: \beditor{\bsnm{{Poletto}}, \binits{G.}},
\beditor{\bsnm{{Suess}}, \binits{S.T.}} (eds.)
\bbtitle{The Sun and the Heliosphere as an Integrated System},
\bsertitle{Astrophysics and Space Science Library}
\bseriesno{317},
\bfpage{147}.
\end{bchapter}
\endbibitem

\bibitem[\protect\citeauthoryear{{Kojima}
  \textit{et~al.}}{2007}]{Kojima2007ASPC}
\begin{bchapter}
\bauthor{\bsnm{{Kojima}}, \binits{M.}},
\bauthor{\bsnm{{Tokumaru}}, \binits{M.}},
\bauthor{\bsnm{{Fujiki}}, \binits{K.}},
\bauthor{\bsnm{{Itoh}}, \binits{H.}},
\bauthor{\bsnm{{Murakami}}, \binits{T.}},
\bauthor{\bsnm{{Hakamada}}, \binits{K.}}:
\byear{2007},
\bctitle{{What Coronal Parameters Determine Solar Wind Speed?}}
In: \beditor{\bsnm{{Shibata}}, \binits{K.}},
\beditor{\bsnm{{Nagata}}, \binits{S.}},
\beditor{\bsnm{{Sakurai}}, \binits{T.}} (eds.)
\bbtitle{New Solar Physics with Solar-B Mission},
\bsertitle{Astronomical Society of the Pacific Conference Series}
\bseriesno{369},
\bfpage{549}.
\end{bchapter}
\endbibitem

\bibitem[\protect\citeauthoryear{{McComas}
  \textit{et~al.}}{2008}]{McComas2008GeoRL}
\begin{barticle}
\bauthor{\bsnm{{McComas}}, \binits{D.J.}},
\bauthor{\bsnm{{Ebert}}, \binits{R.W.}},
\bauthor{\bsnm{{Elliott}}, \binits{H.A.}},
\bauthor{\bsnm{{Goldstein}}, \binits{B.E.}},
\bauthor{\bsnm{{Gosling}}, \binits{J.T.}},
\bauthor{\bsnm{{Schwadron}}, \binits{N.A.}},
\bauthor{\bsnm{{Skoug}}, \binits{R.M.}}:
\byear{2008},
\batitle{{Weaker solar wind from the polar coronal holes and the whole Sun}}.
\bjtitle{\grl}
\bvolume{35},
\bfpage{18103}.
doi:\doiurl{10.1029/2008GL034896}.
\end{barticle}
\endbibitem

\bibitem[\protect\citeauthoryear{{Ohmi} \textit{et~al.}}{2001}]{Ohmi2001JGR}
\begin{barticle}
\bauthor{\bsnm{{Ohmi}}, \binits{T.}},
\bauthor{\bsnm{{Kojima}}, \binits{M.}},
\bauthor{\bsnm{{Yokobe}}, \binits{A.}},
\bauthor{\bsnm{{Tokumaru}}, \binits{M.}},
\bauthor{\bsnm{{Fujiki}}, \binits{K.}},
\bauthor{\bsnm{{Hakamada}}, \binits{K.}}:
\byear{2001},
\batitle{{Polar low-speed solar wind at the solar activity maximum}}.
\bjtitle{\jgr}
\bvolume{106},
\bfpage{24923}\,--\,\blpage{24936}.
doi:\doiurl{10.1029/2001JA900094}.
\end{barticle}
\endbibitem

\bibitem[\protect\citeauthoryear{{Schatten}, {Wilcox}, and
  {Ness}}{1969}]{Schatten1969SoPh}
\begin{barticle}
\bauthor{\bsnm{{Schatten}}, \binits{K.H.}},
\bauthor{\bsnm{{Wilcox}}, \binits{J.M.}},
\bauthor{\bsnm{{Ness}}, \binits{N.F.}}:
\byear{1969},
\batitle{{A model of interplanetary and coronal magnetic fields}}.
\bjtitle{\solphys}
\bvolume{6},
\bfpage{442}\,--\,\blpage{455}.
doi:\doiurl{10.1007/BF00146478}.
\end{barticle}
\endbibitem

\bibitem[\protect\citeauthoryear{{Suzuki}}{2002}]{Suzuki2002ApJ}
\begin{barticle}
\bauthor{\bsnm{{Suzuki}}, \binits{T.K.}}:
\byear{2002},
\batitle{{On the Heating of the Solar Corona and the Acceleration of the
  Low-Speed Solar Wind by Acoustic Waves Generated in the Corona}}.
\bjtitle{\apj}
\bvolume{578},
\bfpage{598}\,--\,\blpage{609}.
doi:\doiurl{10.1086/342347}.
\end{barticle}
\endbibitem

\bibitem[\protect\citeauthoryear{{Suzuki}}{2006}]{Suzuki2006ApJ}
\begin{barticle}
\bauthor{\bsnm{{Suzuki}}, \binits{T.K.}}:
\byear{2006},
\batitle{{Forecasting Solar Wind Speeds}}.
\bjtitle{\apjl}
\bvolume{640},
\bfpage{L75}\,--\,\blpage{L78}.
doi:\doiurl{10.1086/503102}.
\end{barticle}
\endbibitem

\bibitem[\protect\citeauthoryear{{Tokumaru}
  \textit{et~al.}}{2009}]{Tokumaru2009GeoRL}
\begin{barticle}
\bauthor{\bsnm{{Tokumaru}}, \binits{M.}},
\bauthor{\bsnm{{Kojima}}, \binits{M.}},
\bauthor{\bsnm{{Fujiki}}, \binits{K.}},
\bauthor{\bsnm{{Hayashi}}, \binits{K.}}:
\byear{2009},
\batitle{{Non-dipolar solar wind structure observed in the cycle 23/24
  minimum}}.
\bjtitle{\grl}
\bvolume{36},
\bfpage{9101}.
doi:\doiurl{10.1029/2009GL037461}.
\end{barticle}
\endbibitem

\bibitem[\protect\citeauthoryear{{von Steiger} and
  {Zurbuchen}}{2011}]{vonSteiger2011JGRA}
\begin{barticle}
\bauthor{\bsnm{{von Steiger}}, \binits{R.}},
\bauthor{\bsnm{{Zurbuchen}}, \binits{T.H.}}:
\byear{2011},
\batitle{{Polar coronal holes during the past solar cycle: Ulysses
  observations}}.
\bjtitle{Journal of Geophysical Research (Space Physics)}
\bvolume{116},
\bfpage{1105}.
doi:\doiurl{10.1029/2010JA015835}.
\end{barticle}
\endbibitem

\bibitem[\protect\citeauthoryear{{Wang}}{2010}]{Wang2010ApJL}
\begin{barticle}
\bauthor{\bsnm{{Wang}}, \binits{Y.-M.}}:
\byear{2010},
\batitle{{On the Relative Constancy of the Solar Wind Mass Flux at 1 AU}}.
\bjtitle{\apjl}
\bvolume{715},
\bfpage{L121}\,--\,\blpage{L127}.
doi:\doiurl{10.1088/2041-8205/715/2/L121}.
\end{barticle}
\endbibitem

\bibitem[\protect\citeauthoryear{{Wang} and {Sheeley}}{1990}]{Wang1990ApJ}
\begin{barticle}
\bauthor{\bsnm{{Wang}}, \binits{Y.-M.}},
\bauthor{\bsnm{{Sheeley}}, \binits{N.R.} \bsuffix{Jr.}}:
\byear{1990},
\batitle{{Solar wind speed and coronal flux-tube expansion}}.
\bjtitle{\apj}
\bvolume{355},
\bfpage{726}\,--\,\blpage{732}.
doi:\doiurl{10.1086/168805}.
\end{barticle}
\endbibitem

\bibitem[\protect\citeauthoryear{{Wang} and {Sheeley}}{1991}]{Wang1991ApJ}
\begin{barticle}
\bauthor{\bsnm{{Wang}}, \binits{Y.-M.}},
\bauthor{\bsnm{{Sheeley}}, \binits{N.R.} \bsuffix{Jr.}}:
\byear{1991},
\batitle{{Why fast solar wind originates from slowly expanding coronal flux
  tubes}}.
\bjtitle{\apjl}
\bvolume{372},
\bfpage{L45}\,--\,\blpage{L48}.
doi:\doiurl{10.1086/186020}.
\end{barticle}
\endbibitem

\bibitem[\protect\citeauthoryear{{Wang} and {Sheeley}}{2013}]{Wang2013ApJ}
\begin{barticle}
\bauthor{\bsnm{{Wang}}, \binits{Y.-M.}},
\bauthor{\bsnm{{Sheeley}}, \binits{N.R.} \bsuffix{Jr.}}:
\byear{2013},
\batitle{{The Solar Wind and Interplanetary Field during Very Low Amplitude
  Sunspot Cycles}}.
\bjtitle{\apj}
\bvolume{764},
\bfpage{90}.
doi:\doiurl{10.1088/0004-637X/764/1/90}.
\end{barticle}
\endbibitem

\bibitem[\protect\citeauthoryear{{Wang}, {Hawley}, and
  {Sheeley}}{1996}]{Wang1996Sci}
\begin{barticle}
\bauthor{\bsnm{{Wang}}, \binits{Y.-M.}},
\bauthor{\bsnm{{Hawley}}, \binits{S.H.}},
\bauthor{\bsnm{{Sheeley}}, \binits{N.R.} \bsuffix{Jr.}}:
\byear{1996},
\batitle{{The Magnetic Nature of Coronal Holes}}.
\bjtitle{Science}
\bvolume{271},
\bfpage{464}\,--\,\blpage{469}.
doi:\doiurl{10.1126/science.271.5248.464}.
\end{barticle}
\endbibitem

\bibitem[\protect\citeauthoryear{{Wang}, {Robbrecht}, and
  {Sheeley}}{2009}]{Wang2009ApJ}
\begin{barticle}
\bauthor{\bsnm{{Wang}}, \binits{Y.-M.}},
\bauthor{\bsnm{{Robbrecht}}, \binits{E.}},
\bauthor{\bsnm{{Sheeley}}, \binits{N.R.} \bsuffix{Jr.}}:
\byear{2009},
\batitle{{On the Weakening of the Polar Magnetic Fields during Solar Cycle
  23}}.
\bjtitle{\apj}
\bvolume{707},
\bfpage{1372}\,--\,\blpage{1386}.
doi:\doiurl{10.1088/0004-637X/707/2/1372}.
\end{barticle}
\endbibitem

\end{thebibliography}

\clearpage
\end{article} 
\end{document}